\documentclass[preprint, amsmath, amssymb]{revtex4}

 \ifx\pdftexversion\undefined
   \usepackage{graphicx}
 \else
   \usepackage[pdftex]{graphicx}
 \fi

\begin{document}

\title{Acoustic Phonon-Assisted Resonant Tunneling via Single Impurities.}

\author{M.~Gryglas\dag, M.~Baj\dag,\\
B.~Chenaud\ddag, {B.~Jouault}\ddag, 
\\
A.~Cavanna\S~and G.~Faini\S}
\affiliation{\dag~Institute of Experimental Physics, Warsaw University, Poland\\
\ddag~Groupe d'\'Etude des Semiconducteurs, Universit\'e Montpellier II, Montpellier, France\\
\S~Laboratoire de Photonique et de Nanostructures, Marcoussis, France}

\begin{abstract} 

We perform the investigations of the resonant tunneling via
impurities embedded in the AlAs barrier of a single GaAs/AlGaAs 
heterostructure. In the $I(V)$ characteristics
measured at 30mK, the contribution of individual donors is resolved
and the fingerprints of phonon assistance in the tunneling
process are seen. The latter is confirmed by detailed analysis of
the tunneling rates and the modeling of the resonant tunneling
contribution to the current. Moreover, fluctuations of the local
structure of the DOS (LDOS) and Fermi edge singularities are observed.

\end{abstract}


\maketitle

\section{Introduction}

 A considerable amount of resonant tunneling studies was
performed on single barrier GaAs/AlAs/GaAs heterostructures. AlAs
is an indirect gap material with the minimum of the conduction
band at the $X$ point of the Brillouin zone, while in GaAs the minimum
is at the $\Gamma$ point. As one can see in Fig.~1,
the bands are aligned in a way that AlAs
layer forms a barrier for the electrons in the $\Gamma$ valley
(solid line), and a quantum well (QW) for the $X$ valley electrons
(dashed line). There were several
analyses of tunneling through X-valley states including first
observation of negative differential resistance in such
structures~\cite{beresford}, considerations on $\Gamma$-$X$
transfer mechanisms~\cite{teissier}, discussion of momentum
conservation~\cite{finley} as well as through X-minimum-related
donor states (including investigations of the splitting of the ground
state~\cite{itskevich, vitusevich}, pressure coefficients~\cite{gryglas_hp}
and binding energies~\cite{khanin} of donors). However, in those
experiments macroscopic samples were used where many donors
were involved. Recently, interest was more focussed on the observation of
the tunneling through individual objects such as quantum dots or
impurities. In this case it is possible to perform the spectroscopy of
the two dimensional electron gas (2DEG) in the so-called emitter
~\cite{main} as well as of these individual objects which are
involved in the tunneling, e.g. quantum dots ~\cite{lind}.

In this paper we report results of resonant tunneling
experiments performed on GaAs/AlAs/GaAs single barrier junctions in
which single impurities, incorporated in the AlAs layer, are involved.
Some additional features (like Fermi edge singularity and
fluctuations of the LDOS) are also discussed.

\section{Sample fabrication}

Samples were grown by molecular beam epitaxy on a (100)-oriented
Si-doped n-type GaAs wafer ($n_d= 1 \times 10^{18}$ cm$^{-3}$).
The active part of the heterostructure consists of a 10.2nm thick AlAs
barrier incorporating a Si $\delta$-doping in the middle with a concentration
of $n_\delta= 1 \times 10^{10}$  cm$^{-2}$. In some reference samples the 
Si $\delta$-doping is omitted.
The barrier is separated from the heavily 
doped contacts ($n_d= 4 \times 10^{18}$cm$^{-3}$)
by 200nm GaAs spacers in order to obtain high quality
2DEG. Mesa structures of different lateral sizes ranging from 500
$\mu$m down to 100 $nm$ were fabricated. This allows us to have
different numbers of donors within the junction, starting from more
than $10^6$ for the larger mesas to less than 10 for the
smallest ones.

\section{Experimental results for large mesas}

Figure~2 shows current versus voltage characteristics measured at
4.2K on 500$\mu$m mesas, one $\delta$-doped ($10^{10}$cm$^{-2} $,
solid line), the other one undoped (dashed line). In the $I(V)$
curve of the doped mesa we observe an important increase of the
tunneling current at approximately 1 Volt. This is related to the
resonant tunneling through donor states ~\cite{itskevich}, and
since in this case the number of donors is huge ($5\times
10^6$), the observed increase is very broad. The other current
bump at about 2V, present for both samples, is due to the
tunneling through X-minimum quantum well states ~\cite{itskevich,
gryglas_hp}.

\section{Experimental results for small mesas}

If the number of Si impurities inside the barrier decreases~\cite{dellow}
(small mesas), then the contribution of one individual Si impurity can be
resolved 
~\cite{gryglasref_gryglas}. In Fig.~3 we present the $I(V)$
characteristic of a 400nm mesa, measured at very low temperature
(30mK). The diameter is so small that the expected average number
of donors within the mesa is of the order of 10.

We observe several sharp current increases (labeled for clarity), which
are followed either by the decrease of tunneling current ({\it e.g.}
1,3) or by further current increases (2,4). The observed features
can be qualitatively explained in the following way: each of the
increases indicates the beginning of the resonant tunneling
process via an impurity level, whose energy ($\epsilon_d$) becomes
aligned with the Fermi energy ($\epsilon_F$) in the emitter.
Then, as the voltage increases further, the impurity level goes
down with respect to $\epsilon_F$ but remains aligned with
occupied states in the emitter, so that the tunneling current is
maintained. Once it is below the bottom of the subband in the
emitter, the current decreases.

The fact that we observe many steps spread in a relatively wide bias
range indicates that some interdiffusion of Si atoms takes place
during the growth and the Si $\delta$-layer is broadened. The
steps which appear at low biases correspond to donors placed close
to the collector side of AlAs layer.
In electric field, the position at which  the binding energy of the impurity
has its maximum is shifted from the centre of the AlAs layer~\cite{bastard,mailhiot} to the collector side of the structure~\cite{lopez}.

The above considerations lead to the conclusion
that one should
expect a rectangular shape in $I(V)$ characteristics while
measuring the tunneling from 2DEG through a localized state at
very low T. However, a more complex structure is systematically observed.
Let us discuss this point by taking as a reference step 3
(Fig.~\ref{rnmb}). At approximately 825 mV a steep current
increase indicates the beginning of the resonant tunneling, {\it i.e.} 
$\epsilon_d = \epsilon_F$. 
At the current onset, a sharp peak with an overshoot is observed. For
higher values of bias (830-970 mV), the current 
rises systematically but several reproducible peaks with very small amplitudes
are superimposed. Then at approximately 970 mV the tunneling process
is switched off. At this bias 
$\epsilon_d = \epsilon_1$ where $\epsilon_1$ is the bottom of the
subband in the 2DEG (see Fig.~1).


In order to confirm that the maximum of the current observed at the end of the
step (at 970 mV) is related to the switching off 
of the tunneling and thus corresponds to the case where the
impurity level is aligned with the bottom of the 2DEG subband in
the emitter, we further performed magnetotransport measurements. $I(V)$
characteristics have been measured with a magnetic field applied along
the direction of the current. In Fig.~\ref{field} $I(V)$ curves
between 0T and 2.4T are plotted and the shift of the peak related
to the second Landau level is clearly seen as a function of the
magnetic field (the first LL is hidden in the maximum at the end of the
step, but it is revealed in high magnetic fields). 
When $B \rightarrow$ 0T
they converge to $V \approx 970mV$, which is a clear proof that
this voltage corresponds to the bottom of the subband. It means
that the total width of the emitter corresponds to 
140$\pm10$ mV.
Using the correspondence between the Landau level separation
measured in $I(V)$ characteristics and the one calculated in
energy scale $\hbar \omega_c$ (where $\omega_c$ is the cyclotron frequency),
we were able to recalculate
the width of the emitter into energy, obtaining $\epsilon_F$-$\epsilon_1$
$\approx$ 4.8meV.

Referring to Fig.~\ref{field}, we would like to note that within
each $I(V)$ curve a rich structure of very sharp peaks appears in
magnetic field. The average separation between these peaks is of the order of
0.1meV and they are very sensitive to the magnetic
field. Similar observations have already been reported in the literature~\cite{main}. The origin of those structures is not fully understood yet
and some further investigations
must be performed.


When the tunneling process stops, for voltages higher than 970 mV, a
smooth decrease of the tunneling current is observed. It should be
noted that the current never falls to its value observed before
the beginning of the tunneling, but remains at a higher level. We
propose that this feature (current tail), together with the systematic
rise of the current as a function of bias during the tunneling,
is due to the coupling to the environment (contribution of
non-resonant tunneling is negligible in this scale), which is
realized by interactions with phonons~\cite{fujisawa}. This issue will be discussed
in the next paragraphs, where we present our model and results
of the calculations.
In the last part of the paper we will focus on the fine structures
observed within the steps and discuss the current maximum at Fermi energy.

\section{Model}

Tunneling transfer times have been calculated taking into account
only the interactions with the GaAs longitudinal acoustic phonons. Indeed, optical
phonons have much higher energy ($\geq$ 30 meV) and cannot be
responsible for the  observed features that appear at typical
energies of only a few meV. To compute the tunneling time
$\tau_{\mathbf{k}_i}$ from a 2DEG eigenstate $|\mathbf{k}_i>$ to the impurity we use
the Fermi golden rule
\begin{equation}
\frac{1}{\tau_{\mathbf{k}_i}}= \frac{2
\pi}{\hbar}\sum_{\mathbf{q}}\left|<\varphi|{\cal{H_{\mathrm{el-ph}}}}
|\mathbf{k}_i> \right|^2 \delta (\epsilon_f+ \hbar \omega
(\mathbf{q})  - \epsilon_i)
\end{equation}
where $\mathbf{k}_i$ is the wavevector of the electron in the 2DEG plane,
$\epsilon_{i(f)}$ is the energy of the initial (final)
electron state and the kets $|\varphi>$ and $|\mathbf{k}_i>$ refer to
the eigenstates of the impurity and of the 2DEG, respectively. In
this equation the summation is performed over all the final states
with one phonon of energy $\hbar \omega(\mathbf{q})$. Because our
experiments are performed at very low temperature only phonon
emission processes are relevant.

The computation of the matrix elements requires the knowledge of
the wavefunctions. We adopt the envelope function formalism
in the single parabolic band approximation.

The envelope eigenfunctions of the 2DEG are approximated by the
so-called modified Fang-Howard wavefunction which is given
by~\cite{bastard}:

\begin{equation}
<z,\mathbf{r} | \mathbf{k}> = \psi(z) \frac{1}{L} e^{i
\mathbf{k} \cdot
\mathbf{r}}           ~~,\\
\end{equation}
where $\mathbf{r}$ and $z$ are the projections of the electron position vector in the
2DEG plane and in the direction of the current respectively, $L$ is the lateral length and
$\psi(z)$ is defined by
\begin{eqnarray}
\psi(z)=   N z_0 e^{K_b z/2} \mathrm{~~if~} z \le 0~\mathrm{(AlAs~barrier)}, \\ \nonumber
\psi(z)=   N (z+ z_0) e^{-bz/2} \mathrm{~~if~} z \ge 0 ~~\mathrm{(GaAs)}.
\end{eqnarray}
Here $N$, $z_0$ and $b$ are defined by
\begin{eqnarray}
b=
\left[ \frac{48 \pi e^2 m_e}{\kappa \hbar^2}
\left( n_\mathrm{depl} +\frac{11}{32} n_s \right)
\right]^{1/3}~~, \\
N=  \sqrt{\frac{b^3}{2}} \times
\frac{1}
{\sqrt{1+ bz_0 +\frac{1}{2} b^2 z_0^2 (1+ \frac{b}{K_b})}}~~, \\
z_0= \frac{2}
{b+ K_b \frac{m_e}{m_b}}~~,\\
K_b=  2  \frac{ \sqrt{2 m_b (\epsilon_b-\epsilon_i) } }{ \hbar}~~,
\end{eqnarray}
where $m_e$ and $m_b$ are the electron masses in the well and in
the barrier respectively: $m_e=0.067 m_0$, $m_b=0.124 m_0$,
$\kappa$ is the dielectric constant in the well: $\kappa= 12.9
\times \kappa_0$, $n_s$ is the electron density of states: $n_s= 1.32
\times 10^{15}$m$^{-2}$ here, $n_\mathrm{depl}$ is the
concentration of residual acceptors with a typical value of
$6 \times 10^{14}$m$^{-2}$, $m_0$ is the free electron mass and
$\kappa_0$ is the dielectric constant in the vacuum, $\epsilon_b$
is the barrier height and $\epsilon_i$ is the energy of the
electron.

For the sake of simplicity the wavefunction of the impurity inside
the AlAs quantum well is approximated by a spherical gaussian
function:
\begin{equation}
<z,\mathbf{r} |\varphi> = \varphi(x) \varphi(y) \varphi(z-z_i)~~,
\end{equation}
where
\begin{equation}
\varphi(x)=\left( \frac{1}{\pi \sigma^2}\right) ^{\frac{1}{4}}
e^{-\frac{x^2}{2\sigma^2}}~~,
\end{equation}
$z_i$ is the position of the impurity along the growth direction
and $\sigma$ corresponds to the spatial extent of the donor
wavefunction and is comparable to the Bohr radius of a Si impurity.
In the AlAs barrier this value is estimated to be close to 2.6nm. The
adopted form of the donor wavefunction is a very crude
approximation, which cannot give good absolute values for the
tunneling rates. However, we believe that it is good enough for
the qualitative description of the physical phenomena involved.

The electron-phonon interaction is given by:
\begin{equation}
{\cal{H}}_{\mathrm{el-ph}}= \sum_{\mathbf{q}} \left[ \alpha(q)
e^{-i \mathbf{q_\parallel} \cdot \mathbf{r} } e^{-iq_z z} b^{\dagger}_\mathbf{q} +
H.c. \right]~~,
\end{equation}
where $\mathbf{q_\parallel}$ is the phonon wavevector in plane of the 2DEG,
$b^\dagger_{\mathbf{q}}$ is the creation operator for a phonon of
wavevector $\mathbf{q}$ (in all the space directions), and
$\alpha(q)$ are given either by:
\begin{equation}
\alpha(q)= \frac{e e^*_{\mathrm{pz}}}{\kappa}
\sqrt{\frac{ \hbar}{ 2 \rho c_s \Omega}}
\frac{1}{\sqrt{q}}
\end{equation}
for the piezoelectric scattering or by:
\begin{equation}
\alpha(q)= 
D \sqrt{\frac{\hbar}{2 \rho c_s \Omega}} \sqrt{q}
\end{equation}
for the deformation potential scattering, where $D$ is the
deformation potential constant, $\rho$ is the GaAs density, $c_s$
is the sound velocity in GaAs, $e^*_{\mathrm{pz}}$ is the
piezoelectric constant (13eV, 5300kg/m$^3$, 3700m/s and 0.14eV
respectively), $\Omega$ is the volume of the crystal. We have
assumed that the phonons obey the Debye's law. 

The total phonon-assisted tunneling rate $\tau$ is obtained by
summing over the initial states $|\mathbf{k}_i>$:
 $\tau^{-1}= \sum_{\mathbf{k}_i} \tau_{\mathbf{k}_i}^{-1}$
 and an usual calculation~\cite{bastard2} gives:
\begin{equation}
\frac{1}{\tau}= 
\frac{4}{\hbar} \frac{\Omega}{(2\pi)^3} \int_{k_0}^{k_F}\!\!\!\!
dk_i k_i
\int_0^{2 \pi} \!\!\!\!d\theta \int_0^{q_D}\!\!\!\! dq_{\parallel}
q_{\parallel}
| f(q_z) |^2  
|\alpha(q)|^2 \left[ \frac{1}{\hbar c_s} \frac{q}{q_z} \right]
F^2(k_i, q_\|, \theta)~~, \label{eqrate}
\end{equation}
where $k_0= \sqrt{2m_e {\mathrm{max} }(\epsilon_d-\epsilon_1,0)} /
\hbar$, $k_F$ is the 2DEG wavevector at the Fermi energy
$\epsilon_F$, $q_{D}$ is the Debye wavevector, $\theta$
is the angle between $\mathbf{k}_i$ and $\mathbf{q_{\parallel}}$,
$F$ is given by:
\begin{equation}
F=  2 \sqrt{\pi} \sigma \exp \left( -\frac{1}{2}k_i^2
\sigma^2
  -\frac{1}{2}q_{\parallel}^2 \sigma^2 +  k_{i} q_\| \sigma^2
  \cos \theta \right)
\end{equation}
and $f(q_z)$ is a form factor defined by
\begin{equation}
f(q_z)= \int_{-\infty}^{\infty} \!\!\!\! \varphi(z) \psi(z) e^{-i q_z
z} dz \label{formfactor}
\end{equation}
where integration can be performed over the whole space since
$\varphi$ and $\psi$ are localized. Finally $q_z$ is given by the
energy conservation:
\begin{equation}
q_z= \sqrt {
\left( \frac{(\epsilon_i-\epsilon_d)}{\hbar c_s}
\right)^2  - q_\|^2}
\end{equation}
and $q_z^2+q_{\parallel}^2= q^2$.

\section{Results of the calculations}

In this section we present the numerical results.
Figure~\ref{taus} shows the computed phonon-assisted tunneling
rates for both deformation potential and piezoelectric
interactions. The rates are plotted as a function of the donor
energy $\epsilon_d$.
The energy reference is the energy $\epsilon_1$ of the
bottom of the 2DEG subband. Both interactions have the same
qualitative behavior, but the deformation potential interaction
gives a rate one order of magnitude higher than that arising from the piezoelectric
interaction. At the beginning, the rates increase with $\epsilon_d$
to reach a maximum close to $\epsilon_d$=$\epsilon_1$. This
increase is mainly due to the fact that the number of electrons
available for phonon-assisted tunneling increases when
$\epsilon_d$ decreases. Finally, when $\epsilon_d$ diminishes
further and goes below the 2DEG subband, the rates diminish too:
the number of initial states available for tunneling remains
constant but the tunneling rates with phonons of high energy are
strongly reduced by the form factor $f(q_z)$ appearing in
Eq.~\ref{eqrate}. This decreasing tail can be ascribed to the fact
that only acoustic phonons with small $\mathbf{q}$  contribute strongly to the relaxation time in a 2DEG.

In order to describe the observed shape of the tunneling current, one
should take into account not only the phonon-assisted tunneling
but also the elastic one. In the sequential model of the
tunneling, one should consider two processes: transfer from
the emitter to the impurity and from the impurity to the collector.
In our system, the donors observed at low biases are located on the
collector side of the barrier (even if the change of donor binding
energy at the edge of QW is taken into account~\cite{bastard,mailhiot}).
The total current can be expressed by the tunneling rates between the emitter
and the impurity only:
\begin{equation}
I = e \left(
 \frac{1}{\tau}+
\frac{1}{\tau_{\mathrm{df}}}+
\frac{1}{\tau_{\mathrm{pz}}}  \right)
\label{sum}
\end{equation}
where ${\tau_{\mathrm{df}}}$ and ${\tau_{\mathrm{pz}}}$ stand for
the deformation potential and the piezoelectric potential
contributions, respectively, $\tau$ stands for the elastic
tunneling contribution and is inversely proportional to the
square of the overlap between the donor wave function and the
electron wave function in the emitter.

The amplitude of elastic contribution is adjusted to fit the
experimental value of the current measured just above the
threshold, where the contribution of phonon-assisted currents is
still negligible and singularity at  $\epsilon_F$ already vanishes.
If we take
a donor placed close to the collector side of the barrier, e.g.
$z_i$ =7-9nm, we get a reasonable agreement with the experiment
for $\sigma$= 2.2nm-3.2 nm, respectively. The values of $\sigma$
are of the same order of magnitude as the Si Bohr radius. 
The result of the calculation for $z_i$=7nm and $\sigma$= 2.2nm is 
presented in Fig.~\ref{iv} with a solid line.
The calculated 
current rises up to $\epsilon_d= \epsilon_1$, in agreement with the experiment
and the current tails at higher bias are reproduced.
The calculated sharp decrease of the current at
$\epsilon_d=\epsilon_1$ is related to the step-like DOS in the
emitter which is reflected in the elastic contribution.

In order to improve the agreement with the experiment one has to
find the reason why the elastic contribution does not change
so abruptly at $\epsilon_d=\epsilon_1$.
This can happen due to {\it e.g.} the disorder-related broadening of the DOS.
Indeed, in the magnetic field results (see Fig.~\ref{field}) - Landau level N=0 is always
considerably broadened, almost independently of magnetic field.
Possibly there is a band tail at the bottom of the conduction band
induced by disorder. We can take it into account by including in
our model the convolution of the unperturbed DOS with a gaussian.
The result is shown in Fig.~\ref{iv} with a dashed line. The FWHM
of the gaussian was taken to be 0.8meV. One can also expect that
more exact many-body calculations of the current would lead to such
a broadening~\cite{devoret}. Indeed, the Fermi Golden rule is
an approximation that does not take into account tunneling
events with more than one phonon.

\section{The fine structures}

Finally, we want to comment on the detailed structures observed on
the steps, which have not been discussed until now. The first striking
feature is the sharp maximum at the current onset, {\it i.e.}  at the Fermi
energy, see Fig.~\ref{rnmb}. This maximum is due to the Coulomb
interaction between electrons in the 2DEG and the impurity state. This
effect, known in X-ray spectroscopy, was first predicted for the
case of tunneling through a localized level by Matveev and Larkin
~\cite{matveev}. First experimental confirmations were presented
by Geim {\it et al.}~\cite{geim}, who  performed careful
temperature analysis in order to demonstrate a
singular behavior of the current at the Fermi energy. In our case the system changes its
properties due to charging effects during voltage sweep.
Instead of overcoming these effects, we take advantage of
them. In the $I(V)$ curves a hysteresis of the tunneling current is observed: by sweeping up and down the bias,  the same structures are
observed in different bias ranges and they have different widths in the voltage scale.
However, by applying a linear transformation to the bias and a
vertical shift to the current (the need for the latter is related
to charging of some capacitance during the voltage sweep), it is
possible to superimpose the two $I(V)$ curves for both directions
of bias sweep. This is shown in Fig.~\ref{step1} and
Fig.~\ref{step3} where the different axes are labeled for both
bias directions. One clearly sees that the main difference is a shift
of the current onset at the Fermi level in Fig.~\ref{step3}.
It means that the number of electrons in the emitter was slightly
modified. But no matter what the position of the onset was,
the observed maximum remains stuck to the Fermi level.

Following~\cite{matveev} and~\cite{geim}, the Fermi edge
singularity has a power-law shape of the form
\begin{equation}
I \propto \left( \sqrt{(\epsilon_F-\epsilon)^2 + \Gamma_c^2}
\right)^{-\beta} \times \left( \frac{\pi}{2}+ \arctan
\frac{\epsilon_F-\epsilon}{\Gamma_c} \right) \theta(\epsilon_F-
\epsilon)~~, \label{fitfit}
\end{equation}
where $\epsilon$ is the energy of the tunneling electron,
$\hbar$/$\Gamma_c$ is the time an electron can stay in the
impurity state before escaping into the collector side and
\begin{equation}
\beta= \frac{3}{4\pi} \left( k_F d \right)^{-1}~,
\end{equation}
where $d$ is the average distance between the 2DEG and the
impurity state and $k_F$ is the wavevector at the Fermi energy. In
Fig.~\ref{fit} we present an enlargement of step 3 from
Fig.~\ref{step3} for two polarizations. 
Despite the shift at lower biases, the two curves
start to be very similar at biases higher than 0.85 V (in the lower scale of Fig.~\ref{fit}).
It means that at biases higher than 0.85V
the many-body contribution becomes negligible  and the measured current is due
to noninteracting electrons only. The amplitude of this current does not
depend on the bias sweep direction.
This limits the range of acceptable values of fitting parameters.
The voltage corresponding to $\epsilon_F$ was taken from the
experimental results and the value of $k_F = 9.3 \times
10^7$m$^{-1}$ was calculated from the measured
$\epsilon_F$-$\epsilon_1$. Three parameters were fitted, as in
Ref.~\cite{geim}: the amplitude of the many-body current in Eq.~\ref{fitfit} ,
$\Gamma_c$ and $\beta$. The best fit was
obtained for $\Gamma_c$ $\approx$ 0.01 meV and $\beta$ $\approx$
1. The average distance $d$ evaluated from (19) is close to 2.5
nm, which is only about 4 times smaller than expected for impurity
placed at the collector side of the barrier and gives a reasonable order of magnitude. The transfer time $\tau$ is of the order of
100ps.

Another interesting feature that is worth pointing out is the
existence of the hardly resolved peaks within the plateau, in
Fig.~\ref{step3}, between 0.83V up to 0.94V (in the lower scale).
We attribute them to local fluctuations of the DOS in the vicinity of an
impurity ~\cite{lerner, schmidt}. Naturally, we can expect that if the fluctuations are different for some steps, then these steps
are most probably related to different impurities and do not correspond to
ground or excited states belonging to the same impurity. If we
compare step 1 (Fig.~\ref{step1}) and step 3 (Fig.~\ref{step3}),
we see that the detailed characteristics are different, even if 
the signal is much more noisy in step 1, as the values of current are
very small. Thus we argue that we do really observe fluctuations of LDOS
and that step 1 and 3 correspond to two different impurities.

\section{Conclusion}

We have studied resonant tunneling from a 2DEG through the single
impurity state. We see contributions from individual impurities -
each of them appears to be different from the expected rectangular
shape. In particular, there is a current enhancement when the
energy of the impurity approaches the energy of the bottom of the
subband. If the impurity energy decreases further, the current decreases but
does not vanish. We argue that this phenomenon is due to
phonon-assisted tunneling processes. Other additional features like
Fermi edge singularities and fluctuations of the density of states
are clearly resolved. We present an original analysis which
unambiguously demonstrates that the peak at the Fermi level is
related to the Fermi edge singularity and not to the
 fluctuations of the density of states.

\section{Acknowledgements}

We are indebted to X.~Lafosse and L.~Couraud for technical support and to
L. Ferlazzo for the reactive Ion etching. This work
has been partly supported by the Ministry of Scientific
Research and Information Technology (Poland, grant No 2 PO3B 068
24),  by the R\'egion Ile de France (SESAME project 1377) and by the Conseil G\'en\'eral de l'Essonne.

\pagebreak
\begin{figure}
\centering
\includegraphics*[width=0.8\textwidth]{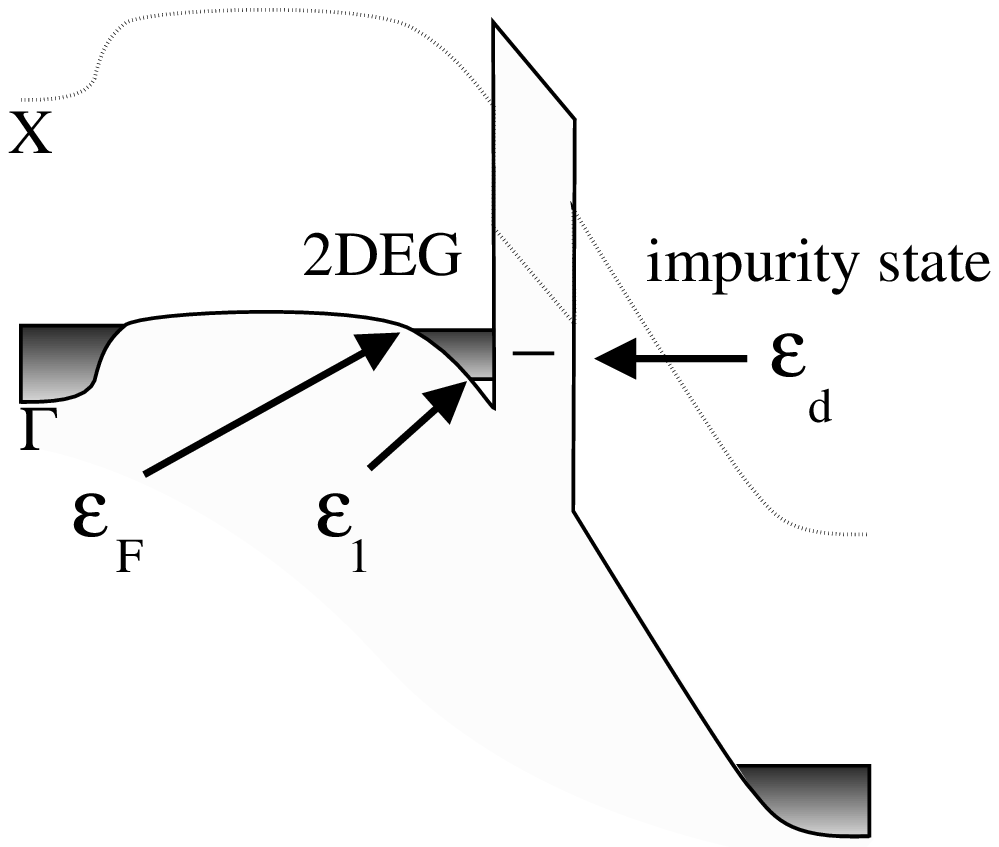}
\caption{Sketch of the conduction band under applied bias. Solid line: $\Gamma$
valley, dotted line: $X$ valley.} \label{sketch}
\end{figure}

\pagebreak
\begin{figure}
\begin{minipage}{0.8\linewidth}
\includegraphics*[width=\textwidth]{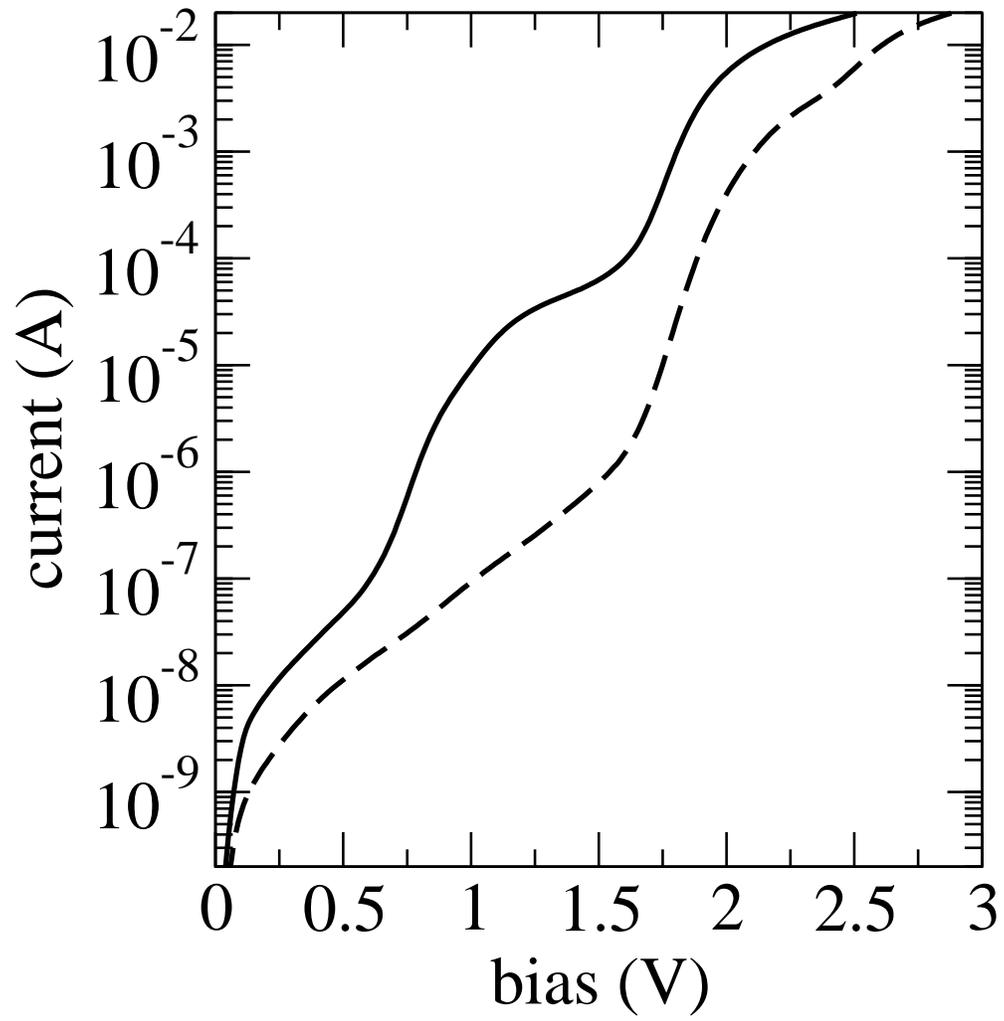}
\caption{$I(V)$ curves for large (0.5mm wide) doped (solid) and
undoped (dashed) mesas.} \label{large}
\end{minipage}
\end{figure}

\pagebreak
\begin{figure}
\begin{center}
\begin{minipage}{0.8\linewidth}%
\includegraphics*[width=\textwidth]{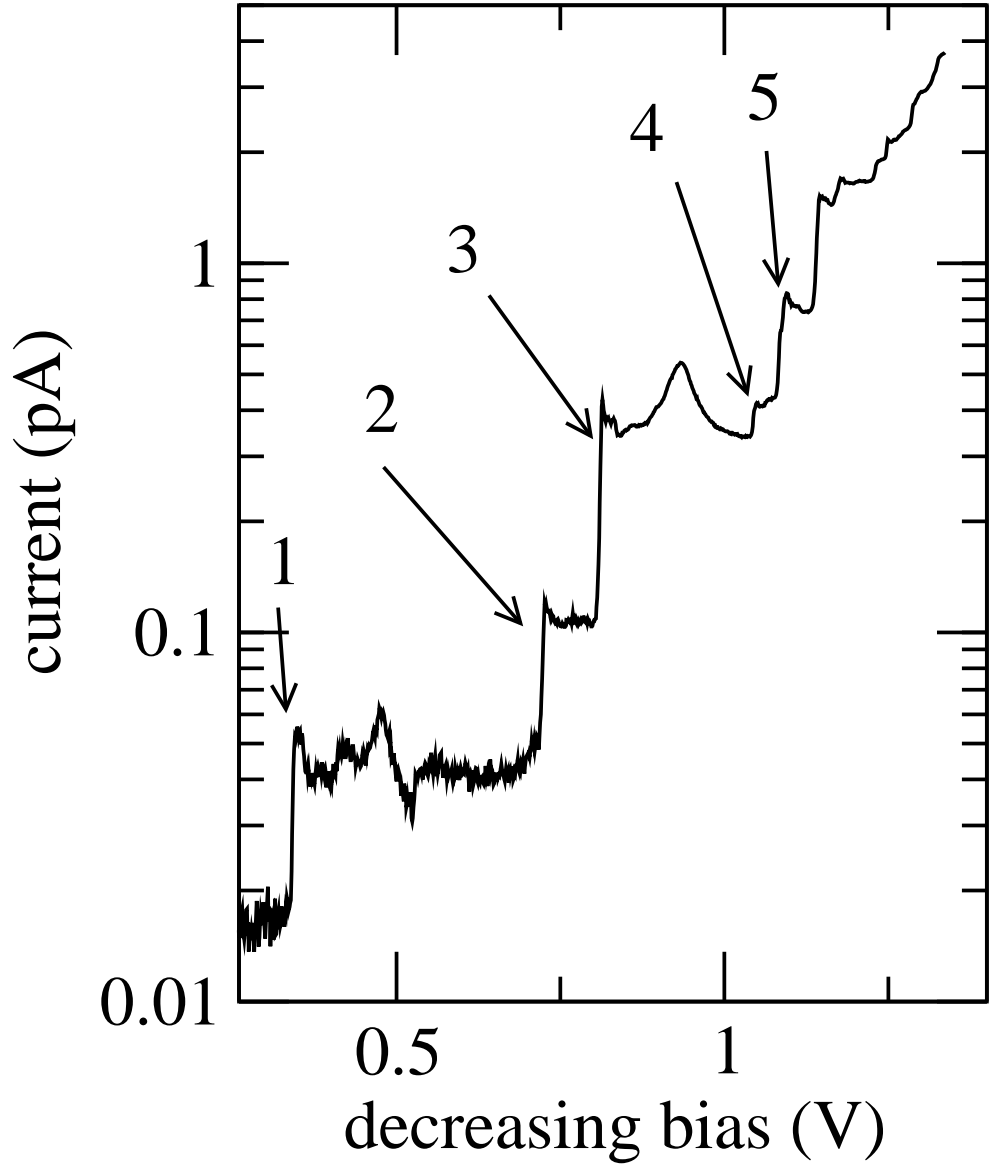}
\caption{$I(V)$ curve for a 400nm mesa measured at very low
temperature (30mK).}
\end{minipage}
\label{rnm}
\end{center}
\end{figure}

\pagebreak
\begin{figure}
\begin{center}
\begin{minipage}{0.8\linewidth}
\includegraphics*[width=\textwidth]{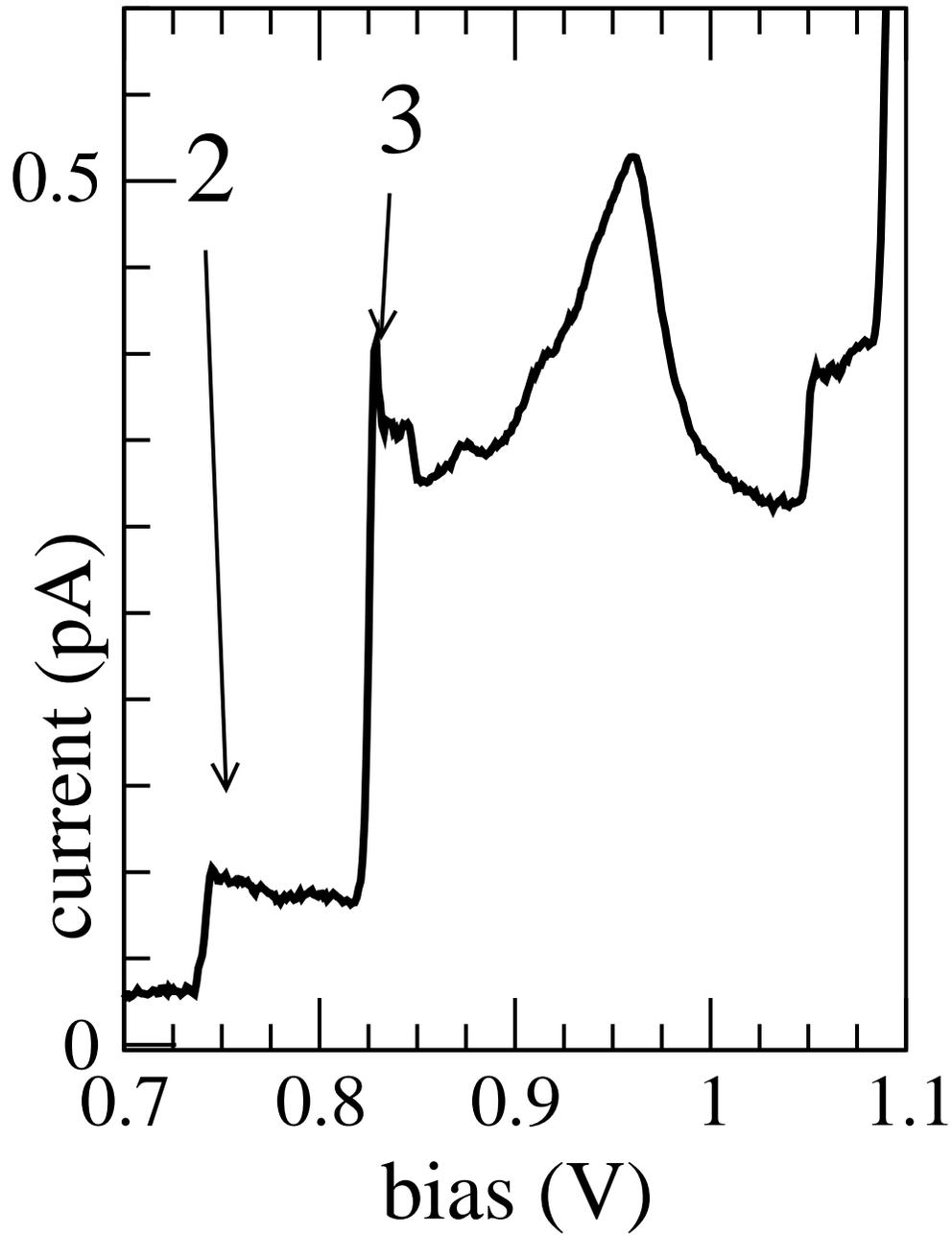}
\end{minipage}
\caption{Detailed measurement of step 3 from Fig.~3.} \label{rnmb}
\end{center}
\end{figure}

\pagebreak
\begin{figure}
\centering
\includegraphics*[width=0.8\textwidth]{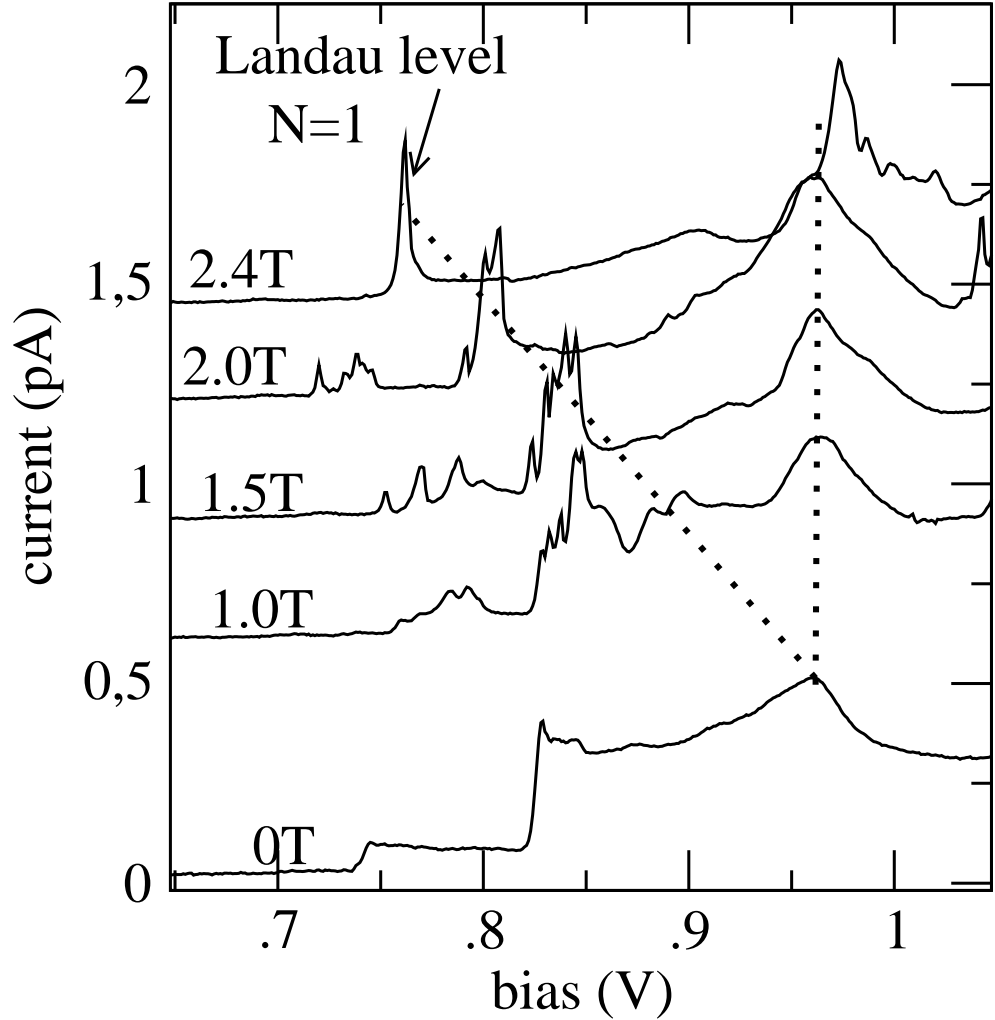}
\caption{$I(V)$ curves at T=30mK in the presence of a magnetic
field of 0, 1, 1.5, 2 and 2.4 T. The field is oriented along the
direction of the current. Curves have been vertically shifted, the
shift is proportional to the magnetic field. Dotted lines are guides
for eyes for the second Landau level and the bottom of the conduction band.} \label{field}
\end{figure}

\pagebreak
\begin{figure}
\begin{minipage}{0.8\textwidth}
\includegraphics*[width=0.9\textwidth]{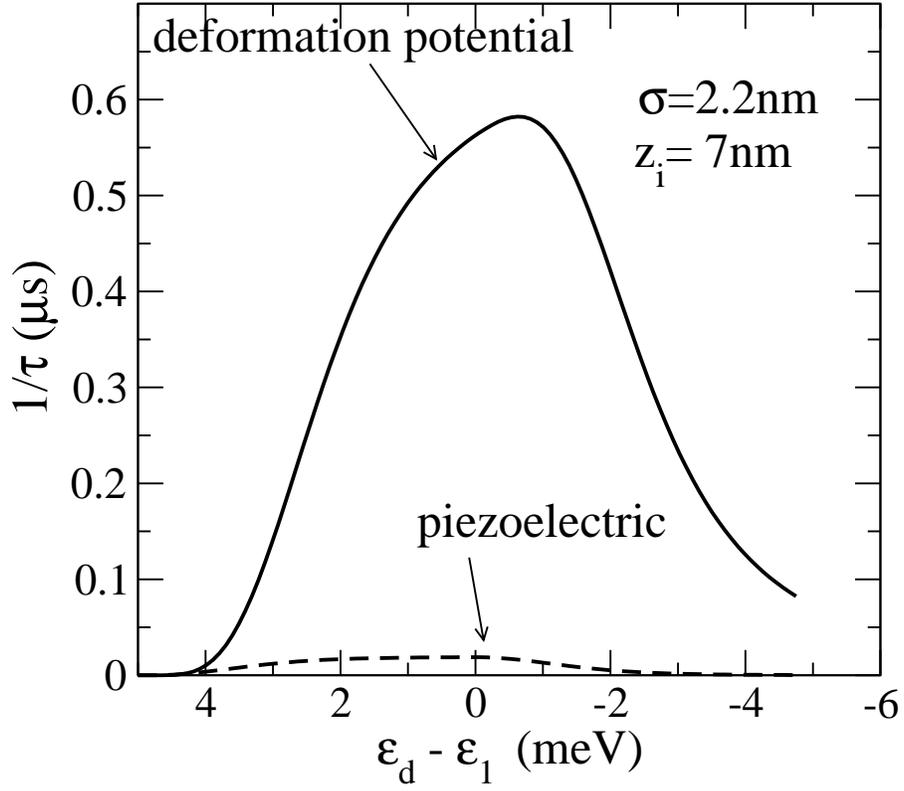}
\caption{Phonon-assisted tunneling rates for both
deformation potential and piezoelectric interactions.}
\label{taus}
\end{minipage}
\end{figure}

\pagebreak
\begin{figure}
\begin{minipage}{0.8\textwidth}
\includegraphics*[width=0.9\textwidth]{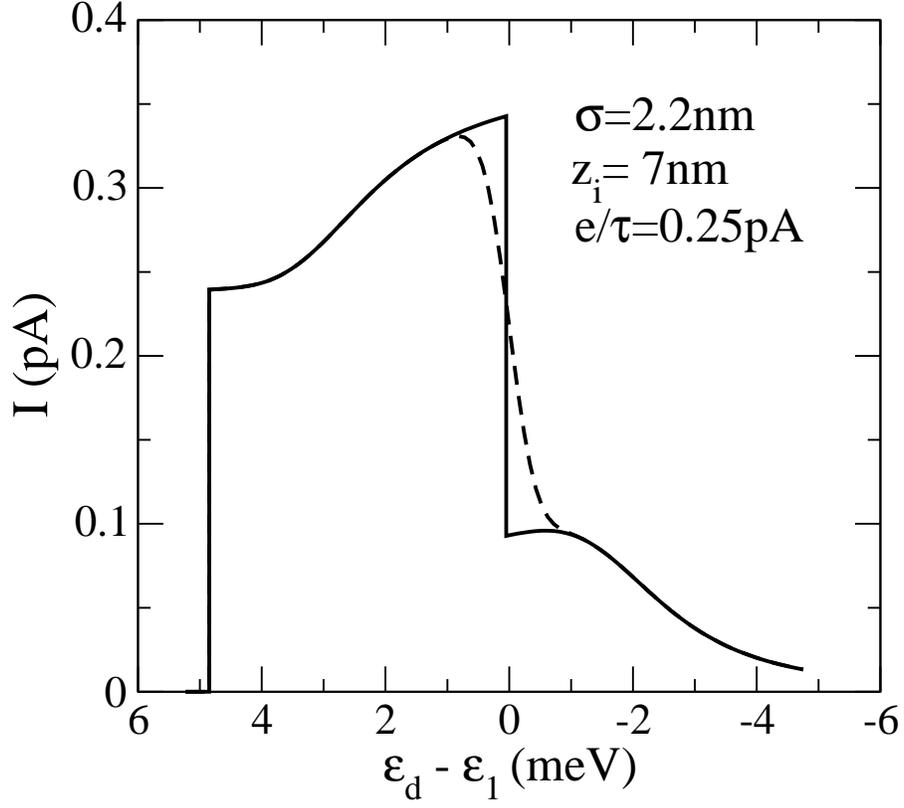}
\caption{Solid line: computed $I(V)$ curve taking into account the elastic
  tunneling contribution and both  phonon-assisted contributions. Dashed
  line: a phenomenological broadening of the emitter density of states has
  been added in the calculation.} \label{iv}
\end{minipage}
\end{figure}

\pagebreak
\begin{figure}
\begin{minipage}{0.8\textwidth}
\includegraphics*[width=0.9\textwidth]{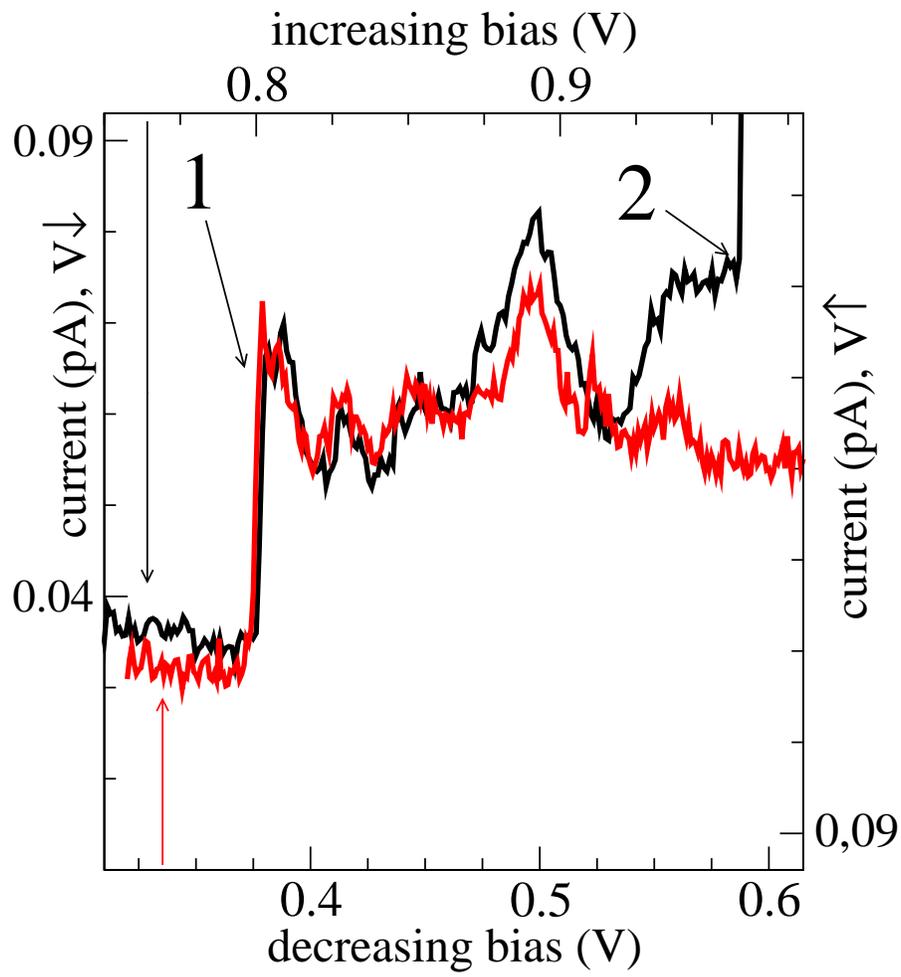}
\caption{First $I(V)$ step for both increasing and
decreasing biases.
}
\label{step1}
\end{minipage}
\end{figure}

\pagebreak
\begin{figure}
\begin{minipage}{0.8\textwidth}
\includegraphics*[width=0.9\textwidth]{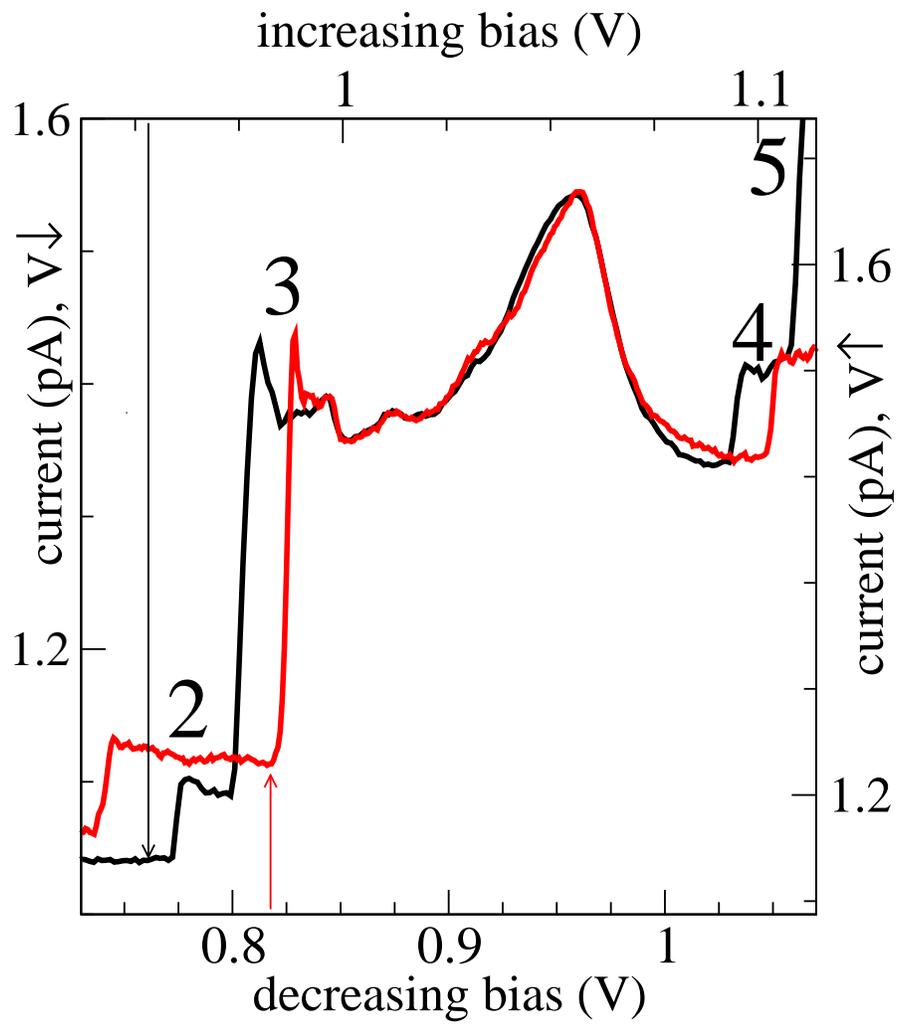}
\caption{Third $I(V)$ step for both increasing and decreasing biases.}
\label{step3}
\end{minipage}
\end{figure}

\pagebreak
\begin{figure}
\centering
\includegraphics*[width=0.7\textwidth]{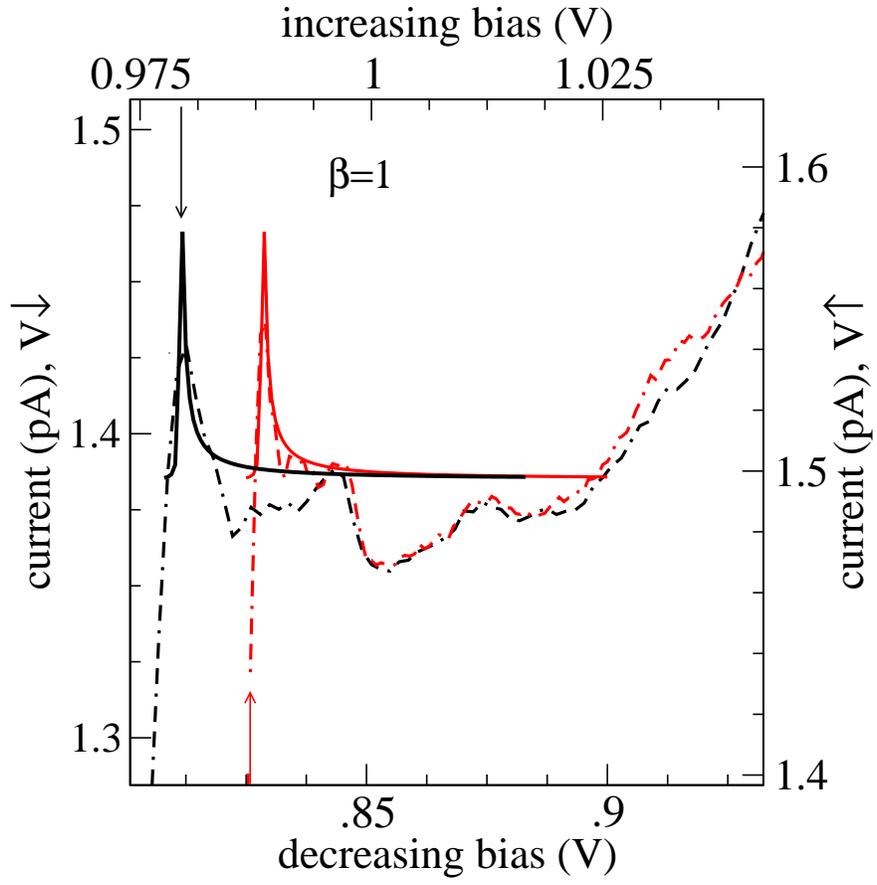}
\caption {$I(V)$ curves for both increasing (black line) and decreasing (gray
  line) biases,
best fit obtained with $\beta$$\approx$1,
$\Gamma_c$$\approx$0.01meV.
} \label{fit}
\end{figure}

\end{document}